\documentclass[aps,prb,showpacs,twocolumn,floats,epsfig]{revtex4}
\usepackage{epsfig,psfrag,subfigure}
\usepackage{graphicx,psfrag,subfigure}
\usepackage{color}
\usepackage{amssymb,amsbsy,amsmath}
\newcommand\beq{\begin{equation}}
\newcommand\eeq{\end{equation}}
\newcommand\bea{\begin{eqnarray}}
\newcommand\eea{\end{eqnarray}}
\newcommand\al{\alpha}

\newcommand\de{\delta}

\newcommand\si{\sigma}
\newcommand\dg{\dagger}

\newcommand\pa{\partial}

\newcommand\non{\nonumber}

\newcommand\ig{\includegraphics}
\newcommand\bib{\bibitem}

\begin{document}

\title{Quenching across quantum critical points in periodic systems:
dependence of scaling laws on periodicity}
\author{Manisha Thakurathi$^1$}
\author{Wade DeGottardi$^2$}
\author{Diptiman Sen$^1$}
\author{Smitha Vishveshwara$^2$}
\affiliation{$^1$Centre for High Energy Physics, Indian Institute of Science,
Bangalore 560012, India \\
$^2$Department of Physics, University of Illinois at Urbana-Champaign, 1110 
W. Green St, Urbana, IL 61801, USA}

\date{\today}

\begin{abstract}
We study the quenching dynamics of a many-body system in one dimension 
described by a Hamiltonian that has spatial periodicity. Specifically, we 
consider a spin-1/2 chain with equal $xx$ and $yy$ couplings
and subject to a periodically varying magnetic 
field in the $\hat z$ direction or, equivalently, a tight-binding model of 
spinless fermions with a periodic local chemical potential, having period $2q$,
where $q$ is a natural number. For a linear quench of the magnetic field 
strength (or potential strength) at rate $1/\tau$ across a quantum critical 
point, we find that the density of defects thereby produced scales as 
$1/\tau^{q/(q+1)}$, deviating from 
the $1/\sqrt{\tau}$ scaling that is ubiquitous to a range of systems. 
We analyze this behavior by mapping the low-energy physics of the system to 
a set of fermionic two-level systems labeled by the lattice momentum 
$k$ undergoing a non-linear quench as well as by performing numerical 
simulations. We also find that if the magnetic field is a superposition of 
different periods, the power law depends only on the smallest period for very 
large values of $\tau$ although it may exhibit a cross-over at intermediate 
values of $\tau$. Finally, for the case where a $zz$ coupling is
also present in the spin chain, or equivalently, where interactions are present
in the fermionic system, we argue that the power associated with the scaling 
law depends on a combination of $q$ and interaction strength. 
\end{abstract}

\maketitle

\section{Introduction}

The quenching dynamics of quantum many-body systems has become a topic
of active research in recent years 
\cite{kibble,zurek,polkov05,calabrese,levitov,dutta,singh,mondal1,
mondal2,polkov08,patane,degrandi,bermudez,vish,pollmann,dziarmaga}. 
In particular, there has been a focus on the effects of slow quenching 
across a quantum critical point (QCP). Here a parameter $\lambda$ in the 
Hamiltonian describing a quantum system is varied linearly (for instance) in 
time at a rate given by 
$1/\tau$, such that the system starts in the ground state far away on one side
of a QCP, crosses the QCP at a critical parameter value $\lambda_c$, and ends 
far away on the other side of the QCP. The final state of the system is in 
fact dominated by the regime close to the QCP\cite{kibble,zurek,polkov05}
where the correlation length of the system diverges as $|\lambda - 
\lambda_c|^{-\nu}$ and the relaxation time (or the inverse of the energy gap 
between the ground state and the first excited state) diverges as $|\lambda 
- \lambda_c|^{-z\nu}$. No matter how small the quenching rate $1/\tau$, the 
relaxation time for the system is larger than $\tau$ because at the QCP, 
there exist modes with energies arbitrarily close to zero. 
The quenching process near $\lambda_c$ is not adiabatic for such modes,
and the system does not reach the ground state of the final Hamiltonian. 
The final state contains a finite density of defects 
which scales as a power of $\tau$ for small $1/\tau$ given by the 
Kibble-Zurek (KZ) form $1/\tau^{d\nu/(z\nu +1)}$. The power thus depends on 
only three quantities; the spatial dimensionality $d$, and critical exponents 
$\nu$ and $z$. For a large class of translationally invariant systems, 
the Hamiltonian decouples into several pairs of momentum modes of 
opposite sign. In such systems, quenching dynamics and power-law scaling 
can be analyzed by mapping the problem to the dynamics of the Landau-Zener 
transition of a two-level system dictated by a time-dependent 
tuning parameter \cite{zurek}. In several such instances in one 
dimension, one has $z=\nu=1$, giving rise to the ubiquitous 
$1/\tau^{1/2}$ scaling of the density of defects scales.

While the above scaling law holds for a range of systems, some
generalizations and deviations thereof are coming to light. For instance, if 
the Hamiltonian is varied across the QCP non-linearly in time, as 
$(t/\tau)^\alpha$, the scaling law becomes $1/\tau^{d\nu \al/ (z\nu \al +1)}$
\cite{mondal2,polkov08}. Effectively, the non-linear quench 
modifies the exponent $\nu$ to $\nu \al$. As another instance, it has also 
been shown that the power law can depend on certain topological features 
of the system \cite{bermudez,vish,wade}. In particular, in Refs. 
\onlinecite{vish} and \onlinecite{wade}, some of the authors of the present 
work have studied quenching dynamics in a two-legged ladder version of the 
Kitaev model. As with the parent Kitaev model on 
the hexagonal lattice \cite{kitaev,feng,nussinov}, the 
two-legged ladder has a large number of sectors (growing exponentially 
with the number of sites) which are distinguished by the eigenvalues ($\pm 1$)
of a number of local $Z_2$-valued conserved quantities \cite{feng,vish}. 
The ladder was mapped to a fermionic $p$-wave superconducting system which has
recently emerged as an exciting topological system in its own right 
\cite{hasan}. It was shown that while quenching through QCPs in certain 
sectors yields the standard $1/\tau^{1/2}$ scaling of the defect density 
(also related to the total probability of excitations), certain periodic 
patterns in the $Z_2$ invariants give rise to a $1/\tau^{2/3}$ scaling. It 
was further conjectured in Ref. \onlinecite{vish} that 
more general scaling laws of the form $1/\tau^{q/(q+1)}$ for the 
slew of integers $q=3,4,\cdots$ may appear in some other sectors, but 
explicit examples of such sectors were not found in that model.

In this work, we show that the presence of a periodically varying parameter 
in an otherwise spatially homogeneous system provides an excellent route for
new and interesting power-law quenching behavior, inclusive of the 
$1/\tau^{q/(q+1)}$ scaling. We demonstrate the role of periodicity in the 
specific case of a one-dimensional spin-1/2 lattice model having homogeneous 
and equal nearest-neighbor $xx$ and $yy$ couplings and subject to a transverse
periodically varying magnetic field having the value $h \cos (Qn + \phi)$
at site $n$. By the Jordan-Wigner transformation, we equivalently 
study a tight-binding system of non-interacting spinless fermions in one 
dimension in which the local chemical potential varies as $h \cos 
(Qn + \phi)$. For $Q = \pi /q$ (where $q=1,2,\cdots$), the system 
exhibits a spatial periodicity of $2q$, fragmenting the Brillouin zone into 
$2q$ regions that are coupled to one another by the periodic potential. In 
contrast to the aforementioned case of pairwise coupling of modes of
opposite momentum, the periodicity presents a more complex structure in 
which a quench
can cause probability amplitudes to shift between these $2q$ modes. While the 
total post-quench excitation probability is still dominated by the QCP 
at $h=0$ and the gapless points at momenta $k = \pm \pi/2$, it depends on 
the intermediate paths taken by the matrix elements connecting the
various fragments of the Brillouin zone; the parameter $\phi$
controls the relative phase between the different paths. 
Using perturbation theory to $q$-th order for small $h$, we explicitly 
illustrate this point for a linear quench where the amplitude $h$ 
varies linearly in time as $t/\tau$ so as to go from $-\infty$ to 
$\infty$. The perturbation provides an effective low-energy Hamiltonian 
for momentum modes around $k = \pm \pi/2$ and reduces the dynamics to the 
generalized Landau-Zener evolution of a set of two-level systems in the 
presence of a $|t/\tau|^q$ time variation, reminiscent of the non-linear 
$t^\alpha$ quench. The behavior of $p_k$, the excitation probability for 
each momentum mode, becomes increasingly complex as the value of $q$ 
increases; however, a simple scaling analysis shows that the total excitation 
probability only gets a contributions from modes lying within a range of $k$ 
given by $\Delta k \sim (\cos (q\phi)/\tau^q)^{1/(q+1)}$ or 
$(\sin (q\phi)/\tau^q)^{1/(q+1)}$, depending on whether $q$ is odd or even,
respectively. Hence the total excitation probability for the 
quench yields the desired $1/\tau^{q/(q+1)}$ scaling law, multiplied
by $(\cos (q\phi))^{1/(q+1)}$ or $(\sin (q\phi))^{1/(q+1)}$. We corroborate
and expand our analytical arguments using numerics. 
It turns out that one can think of the effective low-energy 
theory as either describing a system with $\nu = 1$ and a Hamiltonian varying 
in time with a power $\al = q$, or as a system with $\nu = q$ with a 
Hamiltonian varying linearly in time ($\al =1$). In all cases, 
we obtain a KZ power law of the form $1/\tau^{d\nu \al/(z \nu \al + 1)}$, 
where $d=1$, $z=1$, and $\nu \al = q$.Our results thus show that internal mode 
structure, when combined with critical behavior and
quenching, can give rise to rich out-of-equilibrium dynamics and scaling. 

In another line of investigation, it has been shown that the simple 
post-quench $1/\sqrt{\tau}$ scaling can have drastic modifications due to a 
completely different reason, namely, interactions\cite{degrandi,pollmann}. In 
Ref. \onlinecite{degrandi}, an interacting
system of bosons in the presence of a periodic potential permitting 
one boson per potential minimum was analyzed within a sine-Gordon 
framework. It was shown that a quench in the strength of the periodic 
potential results in the density of defects having a power 
law scaling with an exponent that depends
explicitly on the interaction strength. In the context of our work, 
interactions in a sense are an extreme limit of coupling between momentum 
modes. Borrowing from the analysis of Ref. \onlinecite{degrandi},
we extend our studies of periodic structures to include interactions 
in our fermionic system, or equivalently, to include a $zz$ coupling in the 
spin chain system. We argue that the scaling exponent for the total excitation
probability now depends on both interactions and the periodicity 
$q$ and that this result is also valid for
a generalization of the studies in Ref. \onlinecite{degrandi} 
to the case of $q$ bosons per potential minimum. 

We present our studies as follows. In Sec. II, we introduce the spin 
and fermion models, express the periodic term in the basis of Brillouin 
zone modes and derive the effective low
energy Hamiltonian using perturbation theory. In Sec. III, we analytically 
derive the scaling behavior of the system for a linear quench in $h$ 
and present numerical results of quenching simulations for different values of 
$q$, phase $\phi$ and quenching rate $1/\tau$. To understand the detailed 
dependence of $p_k$ on $k$, we study in the Appendix a two-level problem in 
which the time-dependent term in the Hamiltonian as $|t|^q sgn (t)$ (where 
$sgn$ denotes the signum function). In Sec. IV, we study more complex 
behavior of the periodic potential such as a superposition of two 
commensurate periodic functions; we find that the power law scaling for
the total excitation probability is governed by the smallest period for
very large values of $\tau$ although the power law may show a cross-over at
intermediate values of $\tau$. In Sec. V we consider the effect of 
interactions and we conclude with general remarks in Sec. VI. 


\section{The model and its energy spectrum}

Our starting point is a one-dimensional spin-1/2 model placed in a 
transverse magnetic field whose Hamiltonian is given by
\beq H ~=~ - J ~\sum_{n=-N}^N ~[~ \si_n^x \si_{n+1}^x ~+~ \si_n^y \si_{n+1}^y ~
+~ h_n \si_n^z ~], \label{xxh} \eeq
where $\si_n^\al$ ($\al=x,y,z$) denote the Pauli matrices at site $n$, and we 
are eventually interested in the thermodynamic limit $N \to \infty$.
(We generally set $\hbar$, the exchange coupling $J$ and the 
lattice spacing $a$ equal to unity. When we introduce a quenching time $\tau$,
large or small values of $\tau$ are as compared to $\hbar/J$.)
 Note that $\sum_n \si_n^z$ commutes
with the Hamiltonian. This system can be mapped to a model of 
spinless fermions using the Jordan-Wigner transformation \cite{lieb61}. At 
any site $n$, we map a spin state with $\si_n^z = 1$ or $\si_n^z = -1$ to 
the presence or absence of a spinless fermion at that site; this is
done by introducing a fermion annihilation operator $c_n$ at each site,
and writing the spin at that site as
\bea \si_n^z &=& 2 c_n^\dg c_n -1 = 2 \rho_n - 1 \non \\
\si_n^- &=& \frac{1}{2} (\si_n^x -i \si_n^y)= c_n ~e^{i\pi \sum_{j=-N}^{n-1}
c_j^\dg c_j}, \label{jw} \eea
where $c_j^\dg c_j = 0$ or 1 is the fermion occupation number at site $j$. The
expression for $\si_n^+$ can be obtained by taking the Hermitian conjugate
of $\si_n^-$. The string factor in the definition of $\si_n^-$ is necessary
to ensure the correct anticommutation relations between the fermionic
operators, namely, $\{ c_m, c_n^\dg \} = \de_{mn}$ and $\{ c_m, c_n \} = 0$.

[In this paper, we will use both first quantized notation (wave functions)
and second quantized notation (fermion creation and annihilation operators 
and occupation number basis) as per convenience We 
specify the notation being used where necessary.]

Following the Jordan-Wigner transformation, Eq.~(\ref{xxh}) takes the form
\beq H ~=~ - \sum_{n=-N}^N ~[ 2(c_n^\dg c_{n+1} + c_{n+1}^\dg c_n) ~+~ 2 h_n ~
c_n^\dg c_n], \label{ham1} \eeq
where we have omitted a constant equal to $\sum_n h_n$. The fermionic 
operators can be represented in the momentum basis by the Fourier transform 
\beq c_k ~=~ \frac{1}{\sqrt{2N+1}} ~\sum_{n=-N}^N c_n e^{-ikn}, \eeq
where the momentum $k$ lies in the range 
$[-\pi, \pi]$ and is quantized in units of $2\pi/(2N+1)$;
these operators satisfy the anti-commutation rules $\{ c_k, c_{k'}^\dg \} 
= \delta_{k,k'}$. In momentum space, the first two terms of the Hamiltonian 
in Eq.~(\ref{ham1}) have the tight-binding form
\beq H_0 ~=~ -\sum_{-\pi < k \le \pi} ~4\cos (ka) ~c_k^\dg c_k. \label{ham2} 
\eeq 

As for the last term in Eq.~(\ref{ham1}), we consider the situation where $h_n
= h \cos (Q n + \phi)$. We note that this Hamiltonian appears in the 
Azbel-Hofstadter problem\cite{azbel} of an electron hopping between the sites 
of a square lattice in the presence of a magnetic field applied in the 
perpendicular direction; the magnetic flux per square is proportional to $Q$. 
This problem has been studied extensively, and we will refer to only some of 
the papers here \cite{hsu,sen00}.

For the case $Q=\pi/q$, where $q$ is an integer, the periodicity of the
magnetic field is $2q$. Using the decomposition
\beq h_n ~=~ \frac{h}{2} ~(e^{i(\pi n/q + \phi)} ~+~ e^{-i(\pi n/q + \phi)}),
\eeq
we see that this term couples two fermionic modes with momenta $k_1$ and $k_2$
if $k_1 = k_2 \pm \pi /q$. This periodic term fragments the Hamiltonian into 
a $2q$ dimensional matrix form composed of momentum regions $-\pi + r\pi/q < k
\le -\pi + (r+1)\pi/q$ with $0 \le r \le 2q-1$. The tight-binding term of 
Eq.~(\ref{ham1}) is diagonal in this basis while the periodic potential 
connects neighboring momentum regions. The corresponding matrix elements of 
$H_k$ are given by
\bea && \langle k+r\pi/q | H_k | k+s\pi/q \rangle \non \\
&& = -4 \cos(k+s\pi/q) ~\de_{r,s} - h e^{i\phi} ~\de_{r,s+1} - h e^{-i\phi} ~
\de_{r,s-1}, \non \\
&& \label{hk} \eea
where $0 \le r,s \le 2q-1$, and we have assumed `periodic boundary conditions' 
for the matrix $H_k$, so that $r=-1$ and $2q$ mean $r=2q-1$ and $0$ 
respectively. We have thus reduced Eq.~(\ref{ham1}) to the decoupled form
\beq H ~=~ \sum_{k=-\pi}^{-\pi+\pi/q} ~H_k. \label{ham3} \eeq
Note that the total fermion number ${\cal N} = \sum_{n=-N}^N c_n^\dg c_n
= \sum_{k=-\pi}^\pi c_k^\dg c_k$ commutes with each of the $H_k$ and
is therefore conserved in time.

\begin{figure}[t] \ig[width=3.2in]{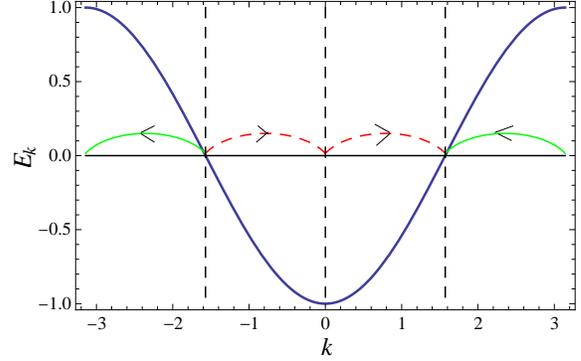}
\caption[]{Picture of intermediate states for the case $q=2$. The zero energy 
states at momenta $k = \pm \pi/2$ are connected to each other through the 
states lying at $k=0$ and $\pi$, as shown by red (dashed) and green (solid) 
lines respectively; the dispersion relation $E_k = - \cos k$ is indicated by a 
thick solid line. The system is described by a set of decoupled Hamiltonians
$H_k$ labeled by $k$ lying in the range $[-\pi,-\pi/2]$; each $H_k$ is a
four-dimensional matrix in which the momenta $k$, $k+\pi/2$, $k+\pi$ and 
$k+3\pi/2$, lying in the regions separated by the vertical dotted lines,
are coupled to each other.} \label{intermed} \end{figure}

We now consider the energy spectrum of $H_k$, focusing on the low energy
excitations that give the dominant contribution to the quench. As will be 
discussed in Sec. III, the initial condition for the quenching dynamics is 
such that the system is at half-filling at all times; we will therefore be 
particularly interested in the states near zero energy which correspond to 
the momenta $k = \pm k_F$, where the Fermi momentum $k_F = \pi/2$. If the 
amplitude $h$ of the magnetic field is zero, the system is gapless and the 
states at $k = \pm \pi/2$ are degenerate with each other. A small, 
non-vanishing value of $h$ (compared with the bandwidth) breaks this 
degeneracy in a manner that we derive using a perturbative expansion in $h$. 
To consider coupling between the $k=-\pi/2$ and $k=\pi/2$ regions, one can go 
through one series of intermediate states lying at $k=-\pi/2 + \pi/q$, $-\pi/2
+ 2\pi/q$, $\cdots$, $\pi/2 - \pi/q$ (with an amplitude equal to $-he^{i\phi}$
at each step), and through another series of intermediate states lying at 
$k=-\pi/2 - \pi/q$, $-\pi/2 - 2\pi/q$, $\cdots$, $\pi/2 + \pi/q$ (with an 
amplitude equal to $-he^{-i\phi}$ at each step). A picture of these two
series of intermediate states is shown in Fig.~\ref{intermed} for $q=2$.
Each of these series consists of $q-1$ intermediate states. At the $q$-th 
order in perturbation theory, we therefore obtain an effective Hamiltonian 
$H_{eff}$ which
has a matrix element between the states at $k = \pm \pi/2$ given by
\bea && \Delta \equiv \langle \pi/2 | H_{eff} | - \pi/2 \rangle ~=~ \langle - 
\pi/2 | H_{eff} | \pi/2 \rangle^* \non \\ 
&& = ~\frac{(-he^{i\phi})^q}{\prod_{s=1}^{q-1} ~(4\cos(-\pi/2 + s\pi/q))} 
\non \\
&& ~+~ \frac{(-he^{-i\phi})^q}{\prod_{s=1}^{q-1} ~(4\cos(-\pi/2 - s\pi/q))},
\label{pert1} \eea 
where the denominators come from factors like $E_{-\pi/2} - E_{-\pi/2 \pm 
s\pi/q} = 4\cos(-\pi/2 + \pm s\pi/q)$ corresponding to the energies in the 
unperturbed Hamiltonian in Eq.~(\ref{ham2}). 

Simplifying Eq.~(\ref{pert1}) gives
\bea && \Delta = \langle \pi/2 | H_{eff} | - \pi/2 \rangle ~=~ \langle - 
\pi/2 | H_{eff} | \pi/2 \rangle^* \non \\
&& = ~(-1)^q h^q ~\frac{e^{iq\phi} + (-1)^{q-1} e^{-iq\phi}}{\prod_{s=1}^{q-1}~ 
(4\sin(\pi s/q))}. \label{pert2} \eea
We thus find that the magnitude of $\Delta$ is given by
\bea |\Delta| &=& \frac{h^q}{4^{q-1}} ~\frac{2 |\cos (q\phi)|}{\prod_{s=1}^{
q-1}~ \sin(\pi s/q)} ~~~{\rm if}~~ q ~{\rm ~is ~odd}, \non \\
&=& \frac{h^q}{4^{q-1}} ~\frac{2 |\sin (q\phi)|}{\prod_{s=1}^{q-1}~ \sin
(\pi s/q)}~~~{\rm if}~~ q ~{\rm ~is ~even}. \label{pert3} \eea
In a sense, the phase $\phi$ governs the relative phase between the two 
paths of intermediate states which connect the low lying states. 
We assume here that $\phi$ is such that $\cos (q\phi) \ne 0$ if $q$ is
odd, and $\sin (q\phi) \ne 0$ if $q$ is even; if these conditions are
violated, we would have to go to higher order perturbation theory to find
a non-zero matrix element connecting the states at $k=\pm \pi/2$.

We can now consider moving slightly away from $k=\pm \pi/2$; then the 
unperturbed energies of the states $k=-\pi/2 + k'$ and $\pi/2 + k'$ are 
given by $-4 k'$ and $4 k'$ respectively. The effective Hamiltonian
describing these two states is then given by the $2 \times 2$ matrix
\beq H_{eff,k'} ~=~ \left( \begin{array}{cc}
- 4 k' & \Delta^* \\
\Delta & 4 k' \end{array} \right), \label{heff} \eeq
where we assume that $\Delta$ continues to be given by the expression in 
Eq.~(\ref{pert2}) because $k'$ is small. The eigenvalues of (\ref{heff})
are given by $\pm \sqrt{(4 k')^2 + | \Delta|^2}$; this is the dispersion 
of a massive relativistic particle whose velocity is equal to the Fermi 
velocity $v_F = 4$ and mass is 
proportional to $|\Delta| \sim h^q$ times $\cos (q\phi)$ or $\sin (q\phi)$.

Hence, $h=0$ corresponds to a QCP where the mass
gap vanishes. Given that the energy vanishes as $|k'|$ if $h=0$ and as $h^q$ 
if $k' = 0$, the dynamical critical exponent and correlation length exponent 
are given by $z=1$ and $\nu = q$, respectively. The correlation length 
exponent $\nu$ thus depends on the periodicity of the magnetic field $h_n$.

\section{Quenching dynamics}

Having established the form of the Hamiltonian for the periodic system and 
its effective low-energy description, we now consider a specific quenching 
protocol.Given the magnetic field $h_n = h \cos (\pi n/q + \phi)$, we vary 
the amplitude of the field in time as $h=t/\tau$, where we refer to $\tau$ as 
the quenching time. At $t=-\infty$ ($h=-\infty$), we start with the ground 
state of the Hamiltonian denoted by $\Psi_0 (-\infty)$. (Throughout this 
section, the symbol $\Psi$ denotes first quantized wave functions). 
$\Psi_0 (-\infty)$ is the state in which 
$\si_n^z = 1$ ($c_n^\dg c_n = 1$) at all the sites where $\cos (\pi n/q + 
\phi) > 0$, and $\si_n^z = -1$ ($c_n^\dg c_n = 0$) at all the sites where 
$\cos (\pi n/q + \phi) < 0$. [If $\cos (\pi n/q + \phi) = 0$ for some value of
$n$, the ground state is not unique in the limit $h=-\infty$, since the states
with $c_n^\dg c_n =0$ and 1 are then degenerate. We therefore assume 
that $\cos (\pi n/q + \phi) \ne 0$ for all values of $n$ and that 
a finite number of values of $\phi$ are avoided. This is
equivalent to the conditions imposed on $\cos (q\phi)$ or $\sin (q\phi)$
after Eq.~(\ref{pert3}).] It is clear that
in the range $1 \le n \le 2q$, $\cos (\pi n/q + \phi)$ is positive for half 
the sites and negative for the other half. Hence, the ground state is 
half-filled in terms of fermions.  We can write the ground state as the 
product of ground states of the $2q$-dimensional Hamiltonians $H_k (h = - 
\infty)$, 
\beq \Psi_0 (-\infty) = \bigotimes \Psi_{0,k} (-\infty), \eeq
where $\Psi_{0,k} (-\infty)$ is a first quantized wave function which is 
obtained as follows. In the limit $h \to - \infty$, the on-site term is much 
larger than the hopping term in Eq.~(\ref{ham1}); hence the Hamiltonian $H_k$ 
becomes independent of $k$ in this limit. This implies that since the ground 
state is half-filled in real space, the states corresponding to the wave 
function $\Psi_{0,k} (t \to -\infty)$ are also half-filled for each value of 
$k$. Thus $\Psi_{0,k} (t \to -\infty)$ denotes the wave function
corresponding to the state in which the $q$ negative energy states of
$H_k (h=-\infty)$ are occupied by fermions and the $q$ positive energy 
states of $H_k (h=-\infty)$ are empty.

The system evolves dynamically according to the time-dependent 
Schr\"odinger equation. The state at time $t$ is once again given by a product
\beq \Psi (t) = \bigotimes \Psi_k (t), \eeq
where the first quantized wave function $\Psi_k (t)$ is obtained by using 
$H_k (t)$ for the time evolution.
In the limit $t \to \infty$ ($h \to \infty$), we reach the state
$\Psi (\infty) = \bigotimes \Psi_k (\infty)$. Since the Hamiltonian
changes time at a finite rate $1/\tau$, we expect that $\Psi (\infty)$
will differ from the ground state of the final Hamiltonian, $\Psi_0 (\infty) 
= \bigotimes \Psi_{0,k} (\infty)$ except in the adiabatic limit. Our goal
is to determine how far $\Psi (\infty)$ is from the final ground state 
$\Psi_0 (\infty)$ as a function of $\tau$, quantifiable by the
total excitation probability $P$. [Note that the final ground 
state $\Psi_0 (\infty)$ is also half-filled; indeed this is the reason for 
choosing $h_n$ to have an even period, $2q$, so that both the initial 
ground state and the final ground state have the same filling. If the period 
of $h_n$ were odd, the initial and final ground states, corresponding to 
$h = -\infty$ and $\infty$, respectively, would have different occupation 
numbers, and it would not be possible for $\Psi_0 (-\infty)$ to evolve in 
time to $\Psi_0 (\infty)$ even in the limit $\tau \to \infty$, since the
dynamics conserves the fermion number].

The formal procedure for evaluating the final state, as employed in our
numerical calculations is as follows.
We define a $2q$-dimensional matrix $M$ whose elements are given by
\beq M_{rs} ~=~ e^{i\phi} ~\de_{r,s+1} ~+~ e^{-i\phi} ~\de_{r,s-1} 
\label{matm} \eeq
for $0 \le s \le 2q-1$, and we again assume `periodic boundary conditions' 
for the matrix $M$.
In the limit $h \to -\infty$, the magnetic term dominates the system and the
eigenvalues of the matrix $H_k$ in Eq.~(\ref{hk}) are the same as those of
$M$, while in the limit of $h \to \infty$, the eigenvalues of $H_k$ are
the same as those of $-M$. Let the $q$ eigenvectors of $M$ corresponding to the
{\it negative} eigenvalues of $M$ be denoted by $\psi_i$, where $i=1,2,\cdots,
q$; we take the $\psi_i$ to form an orthonormal set. For $h \to - \infty$, the 
ground state is one in which the $q$ states corresponding to the wave
functions $\psi_i$ are occupied and the remaining $q$ states (corresponding to
the positive eigenvalues of $M$) are unoccupied. Next, let us assume that 
under time evolution from $t= -\infty$ to $+\infty$ using $H_k (t)$, the $q$ 
wave functions $\psi_i$ evolve into $\psi_{i,k} (\infty)$. Namely,
\beq \psi_{i,k} (\infty) ~=~ {\cal T} ~\exp \left(-i \int_{-\infty}^\infty ~dt~
H_k (t) \right) ~\psi_i, \label{psiik} \eeq
where $\cal T$ denotes the time-ordering symbol which is required because 
$H_k (t)$ varies in time since $h = t/\tau$. At $t=\infty$, the ground state 
of $H_k$ is one in which the states corresponding to the wave functions 
$\psi_i$ are {\it excited} states and are therefore unoccupied, while the 
states corresponding to the remaining eigenstates of $M$ are occupied. The 
probability of being in an excited state at $t=\infty$ is then given by
\bea p_k ~=~ \sum_{i=1}^q ~\sum_{j=1}^q ~|\langle \psi_i | \psi_{j,k} (\infty)
\rangle|^2. \label{pk} \eea
(With this definition, $p_k$ always lies in the range $[0,q]$). Finally
we obtain the total excitation probability $P$ by integrating over $k$, namely,
\beq P ~=~ \int_{-\pi}^{-\pi+\pi/q} ~\frac{dk}{\pi} ~p_k. \label{prob} \eeq
This is a measure of how far $\Psi (\infty)$ is from the ground state $\Psi_0
(\infty)$ at $t=\infty$; if $P=0$, $\Psi (\infty) = \Psi_0 (\infty)$ up to a 
phase. (We have defined the normalization in Eq.~(\ref{prob}) in such a way 
that 
$P=1$ if $p_k = q$ for all values of $k$). In terms of the spins $\si_n^z$ at 
different sites, $P$ tells us the number of spins per site which point in the 
wrong direction at $t=\infty$, i.e., in the direction opposite to that given 
by the energetically favorable state for Eq.~(\ref{ham1}) in the limit $h = 
\infty$. Thus the total excitation probability $P$ is related to the density 
of defects in the final state, i.e., density of spins pointing in the wrong 
direction. 


{\bf Analytical treatment:-} We first capture the broad 
features of the quenching dynamics analytically by focusing on small $h$ before 
delving into a more detailed numerical analysis. Although the quench tunes 
from $h=-\infty$ to $h=\infty$, the excitations are mainly produced during 
the time when $h$ is close to zero, i.e., 
when $|h|$ is much smaller than the band width. This is because 
$h=0$ corresponds to a QCP where there are states lying arbitrarily close to 
zero energy. No matter how large the quenching time $\tau$, there are
states whose energy is less than $1/\tau$; these are the states for which
the dynamics is not adiabatic, and hence contributing significantly to the 
excitation probability. 

We thus consider the quenching problem in the basis of the effective 
Hamiltonian given in Eq.~(\ref{heff}) for the two states at $k=-\pi/2 + k'$ 
and $\pi/2 + k'$ obtained by perturbation for small $h$. As pointed out 
earlier, $|\Delta|$ scales as $h^q$ times $\cos (q\phi)$ or 
$\sin (q\phi)$ depending on
whether $q$ is odd or even. Now, if $h$ is varied in time as $t/\tau$, the 
time-dependent Schr\"odinger equation for the first quantized wave functions 
$u_{1k}$ and $u_{2k}$ of each pair of momentum states takes the form 
\beq i \frac{d}{dt} \left( \begin{array}{c}
u_{1k} \\
u_{2k} \end{array} \right) = \left( \begin{array}{cc}
- 4 k' & (t/\tau)^q f(q\phi) \\
(t/\tau)^q f^*(q\phi) & 4 k' \end{array} \right)~ \left( \begin{array}{c}
u_{1k} \\
u_{2k} \end{array} \right), \label{tds1} \eeq
where $f(q\phi)$ is equal to $\cos (\phi)$ or $\sin (\phi)$ times factors 
given in Eq.~(\ref{pert3}) which are independent of $\tau$ and $\phi$. 
We can assume that $f(q\phi)$ is real and positive by gauging away any
phase dependence by a relative phase rotation between $u_{1k}$ and 
$u_{2k}$. Multiplying Eq.~(\ref{tds1}) by $(\tau^q /f(q\phi))^{1/(q+1)}$ and 
defining $t'=t(f(q\phi)/\tau^q)^{1/(q+1)}$, we obtain
\bea && i \frac{d}{dt'} \left( \begin{array}{c}
u_{1k} \\
u_{2k} \end{array} \right) \non \\
&& = \left( \begin{array}{cc} 
- 4 k' \left( \frac{\tau^q}{f(q\phi)} \right)^{1/(q+1)} & t'^q \\
t'^q & 4 k' \left( \frac{\tau^q}{f(q\phi)} \right)^{1/(q+1)} \end{array} 
\right) \left( \begin{array}{c}
u_{1k} \\
u_{2k} \end{array} \right). \non \\
&& \label{tds2} \eea
As a specific case, let us assume that $q$ is odd; then $t'^q \to \pm \infty$ 
for $t' \to \pm \infty$ respectively. The ground state of the $2 \times 2$ 
Hamiltonian in Eq.~(\ref{tds2}) is then given by $\psi_- = (1/\sqrt{2}) 
(1,1)^T$ for $t' \to - \infty$ and by $\psi_+ = (1/\sqrt{2}) (1,-1)^T$ for 
$t' \to \infty$ (here $T$ denotes the transpose of a row vector).
It is now clear that if we start in the state $\psi_-$ at $t'=-\infty$, the
excitation probability $p_k = |(u_{1k} (t'), u_{2k} (t'))^* \psi_+|^2$ 
at $t'=\infty$ must be a function of a single variable given by $k' (\tau^q /
f(q\phi))^{1/(q+1)}$. Furthermore, by general quantum mechanical arguments,
the excitation probability $p_k$ must be zero for $\tau \to \infty$ 
(adiabatic limit) and 1 for $\tau \to 0$ (limit of sudden change in the 
Hamiltonian). Hence, if $\tau$ is large, $p_k$ gains a significant contribution
only from values of $k'$ for which $|k'| (\tau^q /f(q\phi))^{1/(q+1)} \lesssim 
1$, i.e., $|k'| \lesssim (f(q\phi)/\tau^q)^{1/(q+1)}$. The exact form of $p_k$
depends on various parameters and has characteristic oscillations which
are analyzed in the Appendix as well as in the numerics below.

It now follows that the total excitation probability is given by
\bea P &=& \int_{-\pi/2 - \pi/(2q)}^{-\pi/2 + \pi/(2q)} ~\frac{dk}{\pi} ~p_k 
\non \\
&\simeq& \int_{-\infty}^{\infty} ~\frac{dk'}{\pi} ~p \left[ k' \left(\frac{
\tau^q}{f(q\phi)} \right)^{1/(q+1)} \right] \non \\
&\sim& \left( \frac{f(q\phi)}{\tau^q} \right)^{1/(q+1)}. \label{kz} \eea
The second equation in Eq.~(\ref{kz}) follows if $\tau$ is large enough
so that the range of $k'$, which is of the order of 
$(f(q\phi)/\tau^q)^{1/(q+1)}$, is much less than the width of the momentum 
band, $\pi/q$; this justifies changing the limits of the integral in 
(\ref{kz}) from $[-\pi/2 - \pi/(2q),-\pi/2 +\pi/(2q)]$ to $[-\infty,\infty]$. 
The third equation in (\ref{kz}) follows because $p_k$ is very small unless 
$k'$ lies in a range of the order of $(f(q\phi)/\tau^q)^{1/(q+1)}$.
We thus see that $\tau$ should be much larger than $\hbar /J$ for the 
power-law scaling in the last equation in (\ref{kz}) to hold.

{\bf Results:-} Equation (\ref{kz}) is our central result. 
Contained in it is the $1/\tau^{q/(q+1)}$ scaling behavior of the total 
excitations probability arising from spatial periodicity. The dependence on 
$q\phi$ shows that the original off-set in the potential $h_n = h \cos 
(\pi n/q + \phi)$ controls
the amount of mixing between the various states, as argued in the
perturbative derivation of the effective Hamiltonian of Eq.~(\ref{heff}). 
While this off-set does not contribute to the scaling form, it determines the 
magnitude of the probability amplitude that shifts between the 
different states due to the quench.

{\bf Numerical treatment:-} We now test our results for various values of 
$\tau$, $q$ and $\phi$ and study the detailed behavior of the quenching 
dynamics through numerical simulations. The procedure has been described
above. Briefly, we begin with a state $\psi_i$ corresponding to one of the 
negative eigenvalues of the matrix $M$ in Eq.~(\ref{matm}). We then use $H_k
(t)$ given in Eq.~(\ref{hk}), with $h = t/\tau$, to evolve $\psi_i$ as shown 
in Eq.~(\ref{psiik}) to find the state $\psi_{i,k} (\infty)$. After repeating 
this calculation for all $i$ from 1 to $q$, we calculate the excitation 
probability $p_k$ following Eq.~(\ref{pk}). We then integrate over $k$ 
as in Eq.~(\ref{prob}) to obtain the total excitation probability $P$.

One remark on the scaling forms for odd versus even values of $q$ is
in order here. As made explicit in Eq.~(\ref{ham3}), the Hamiltonian
$H_k$ in Eq.~(\ref{hk}) connects together all values of $k$ which are 
separated by multiples of $\pi/q$, and $k$ lies in the range $[-\pi,-\pi+ 
\pi/q]$. This is the range of $k$ considered for our numerical calculations.
For the low-lying states, the value of $k$ which lies in the above range and 
which is also connected to $-\pi/2$ is given by $k_0=-\pi$ if $q$ is even and 
$k_0= -\pi+\pi/(2q)$ if $q$ is odd. We therefore work with the scaling variable
$(k-k_0) (\tau^q/\sin(q\phi))^{1/(q+1)}$ if $q$ is even and $(k-k_0)(\tau^q/
\cos(q\phi))^{1/(q+1)}$ if $q$ is odd.

For $q=1$, we have $h_n = (t/\tau) cos (\pi n)$. This problem can be
solved analytically as discussed in Ref. \onlinecite{zurek}. For $h_n
\cos (\pi n)$, there is a direct matrix element between states at momenta
$k$ and $k+\pi$. Hence, for every value of $k$ lying in the range $[-\pi,0]$, 
we have a two-level system for which an exact expression for the excitation 
probability can be found using the Landau-Zener treatment\cite{lz}. We find 
that $p_k = e^{-2\pi \tau \sin^2 k}$ and the total excitation probability is
\beq P_{q=1} ~=~ \int_0^\pi ~\frac{dk}{\pi} ~p_k ~\sim~ \frac{1}{\sqrt{\tau}} 
\eeq 
if $\tau \gg 1$. 

\begin{figure}[t] \ig[width=3.2in]{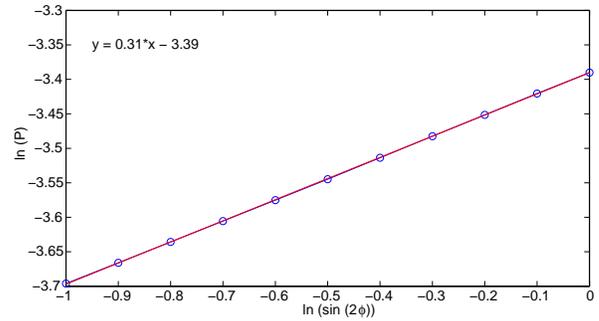}
\caption[]{Logarithmic plot of $P$ versus $\sin (2\phi)$ for $q=2$ and
$\tau = 8$. The linear fit to the points has a slope of $0.31$.} 
\label{sin2p} \end{figure}

For $q=2$, the arguments above indicate that $P$ should scale as 
$(\sin(2\phi))^{1/3}/\tau^{2/3}$. For the special case $\phi = \pi/4$, the 
dependence of $P$ on $\tau$ was studied in Ref. \onlinecite{vish}; numerical
studies confirmed the $-2/3$ power law. We now present the dependence of $P$
on $\sin(2\phi)$ in Fig.~\ref{sin2p}; a linear fit to the logarithmic plot 
gives a slope of $0.31$ which is close to a $1/3$ power law. The fact that $P$
exhibits a power-law scaling with respect to both $\tau$ and $\sin (2\phi)$ 
(for $q=2$) confirms the validity of the perturbation theory developed in 
Eqs.~(\ref{pert1}-\ref{pert3}) for $h$ close to zero and the fact that the 
excitation probability is dominated by the behavior for small $h$.

\begin{figure}[t] \ig[width=3.2in]{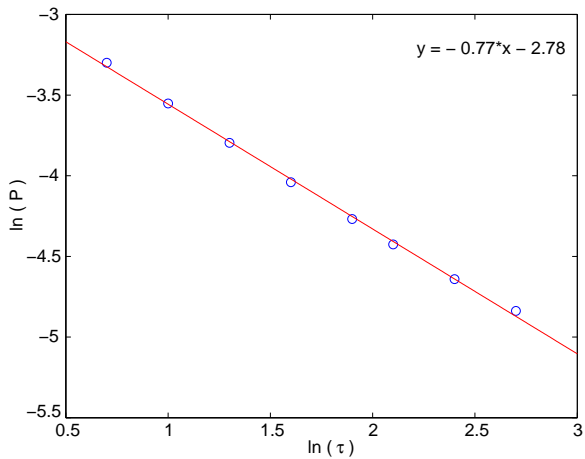}
\caption[]{Logarithmic plot of $P$ versus $\tau$ for $q=3$ and $\phi = 0$. The
linear fit to the points has a slope of $-0.77$.} \label{Pq3tau} \end{figure}

\begin{figure}[t] \ig[width=3.2in]{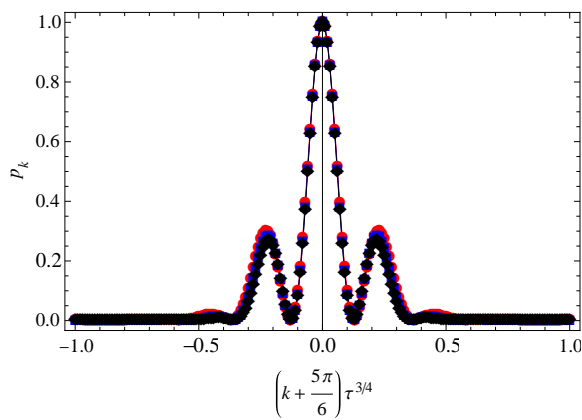}
\caption[]{Plot of $p_k$ versus $(k+5\pi/6)\tau^{3/4}$ for $\tau = 2$ (red), 
4 (blue) and 8 (black), for $q=3$ and $\phi = 0$.} \label{pkq3} \end{figure}

For $q=3$, our arguments show that $P$ should scale as 
$(\cos(3\phi))^{1/4}/\tau^{3/4}$. Fig.~\ref{Pq3tau} shows a logarithmic plot 
of $P$ versus $\tau$ for $\phi = 0$; a linear fit gives a slope of $-0.77$ 
which is in good agreement with the expected value of $-3/4$. Figure 
\ref{pkq3} shows a plot of $p_k$ versus $(k + 5\pi/6) \tau^{3/4}$ for three 
values of $\tau$, for $\phi = 0$; the three curves are seen to coincide, 
indicating that $p_k$ is indeed a function of the scaling variable $(k + 
5\pi/6) \tau^{3/4}$ as indicated in Eq.~(\ref{tds2}), and only a range of 
values of $k + 5\pi/6$ of the order of $1/\tau^{3/4}$ is seen to contribute 
significantly to the total excitation probability. 

\begin{figure}[t] \ig[width=3.2in]{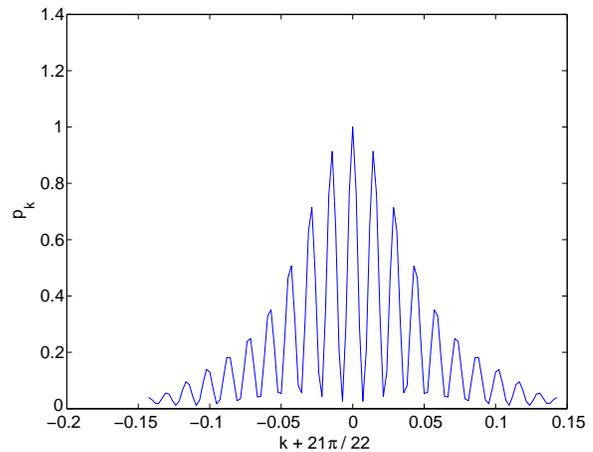}
\caption[]{Plot of $p_k$ versus $k+21\pi/22$ for $q=11$, $\phi = 0$, and
$\tau =16$.} \label{pkq11} \end{figure}

Figure \ref{pkq3} shows several oscillations in $p_k$ as a function of $k$. To
obtain a clearer understanding of these oscillations, we study in the Appendix 
a two-level system in which the one of the terms in the Hamiltonians varies 
in time as $|t|^q sgn(t)$; this is motivated by the form of the Hamiltonian 
in Eq.~(\ref{tds2}). We find that the number of oscillations increases 
with $q$. Our analysis also provides a derivation of the period of oscillation
for large $q$, indicating that the oscillations in $k$ have a period that also
scales as $1/\tau^{q/(q+1)}$; this follows from the statement proved in the 
Appendix that, for large values of $q$, the oscillations have a period $\pi /2$
in a parameter $b$ which is equal to the quantity $4 k' (\tau^q /f(q\phi))^{1/
(q+1)}$ in Eq.~(\ref{tds2}). 
To see the manner in which $p_k$ depends on $q$, we have plotted $p_k$
versus $k+21\pi/22$ for $q=11$ and $\phi = 0$ in Fig.~\ref{pkq11}. Note that 
for $q=11$, $p_k$ is expected to be a function of $(k+21\pi/22) \tau^{11/12}$;
the figure indicates that only a range of values of $k+21\pi/22$ of the order 
of $1/\tau^{11/12}$ contributes substantially to the excitation probability. 
The number of oscillations in $p_k$ shows a clear increase compared 
to the case of $q=3$ in Fig.~\ref{pkq3}.

Our numerical simulations thus confirm the scaling behavior of $P$ as a 
function of the quenching rate and off-set $\phi$, showing that the scaling 
holds for a significant range of parameter space. They also illustrate the 
exact manner in which $p_k$ varies as a function of momentum $k$ and 
periodicity $q$. 

\section{More complex periodic structures}

In the previous sections, we have considered a magnetic field given by
$h_n = h \cos (\pi n/q + \phi)$ assuming even periodicity $2q$, and we have 
varied $h$ in time as $t/\tau$. Our analysis of the energy spectrum close to 
the points $k=\pm \pi/2$ and the scaling of the excitation probability were
based on perturbation theory. We can now ask if other periodic forms for 
$h_n$ lead to similar results. 

We again assume that $h_n = h \cos (Qn + \phi)$, but we set $Q = \pi p/q$,
where $Q < 2 \pi$, $p$ and $q$ are relatively prime integers, and $p$ is odd.
These conditions ensure that the period of $h_n$ is an even integer given by 
$2q$, so that the ground states for $h$ very large and either positive or 
negative are both half-filled. Once again we can do perturbation theory to 
obtain the matrix element between the states at $k=\pm \pi/2$; we find that 
at the lowest order in $h$ which is given by $h^q$, the magnitude of the 
matrix element is given by Eq.~(\ref{pert3}) {\it regardless} of the value of 
$p$. This follows from the number theoretic
fact that the sets of integers $(1,2,\cdots,q-1)$ and $(p,2p,\cdots,p(q-1))$
modulo $q$ are identical if $p$ and $q$ are relatively prime \cite{niven};
thus the set of positive numbers $|\sin (\pi sp/q)|$, for $s=1,2,\cdots,q-1$,
is the same for all values of $p$ which are relatively prime to $q$.
Hence the effective two-level problem given in Eq.~(\ref{tds1}) depends
only on $q$ and not on $p$.

\begin{figure}[t] \ig[width=3.2in]{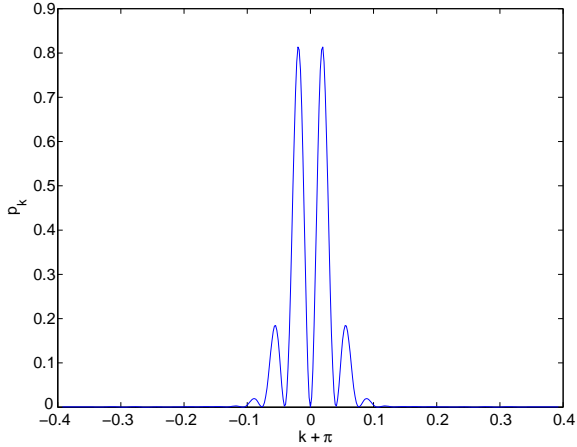}
\caption[]{Plot of $p_k$ versus $k+\pi$ for $Q=\pi/4$, $\phi = \pi/8$, and
$\tau =8$.} \label{pkqpi4} \end{figure}

\begin{figure}[t] \ig[width=3.2in]{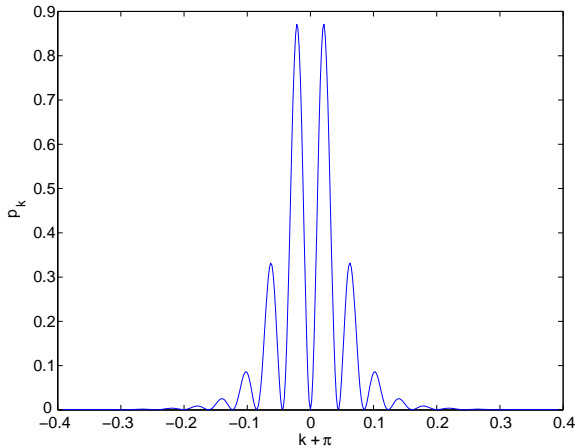}
\caption[]{Plot of $p_k$ versus $k+\pi$ for $Q=3\pi/4$, $\phi = \pi/8$, and
$\tau =8$.} \label{pkq3pi4} \end{figure}

As a simple example, we compare the cases given by $Q=\pi/4$ and $Q=
3\pi/4$. For both of these, $h_n$ has period 8, and for small $h$, 
perturbation theory leads to the same expression for the magnitude of the 
matrix element given in Eq.~(\ref{pert3}) between the states at $k=\pm \pi/2$,
namely, $|\Delta| = (1/16) h^4 |\sin (4 \phi)|$; we therefore obtain the 
same effective two-level problem given in Eq.~(\ref{tds1}). However, the 
complete 8-dimensional Hamiltonians $H_k$ are not 
identical for $Q=\pi/4$ and $3\pi/4$ for arbitrary values of $h$ and $k$, 
even if we allow for unitary transformations. It is therefore interesting to
compare the results for the excitation probabilities $p_k$ in these two
cases. This is shown in Figs.~\ref{pkqpi4} and \ref{pkq3pi4}; we see that the
forms of $p_k$ are qualitatively very similar although they differ 
quantitatively in the two cases.

\begin{figure}[t] \ig[width=3.2in]{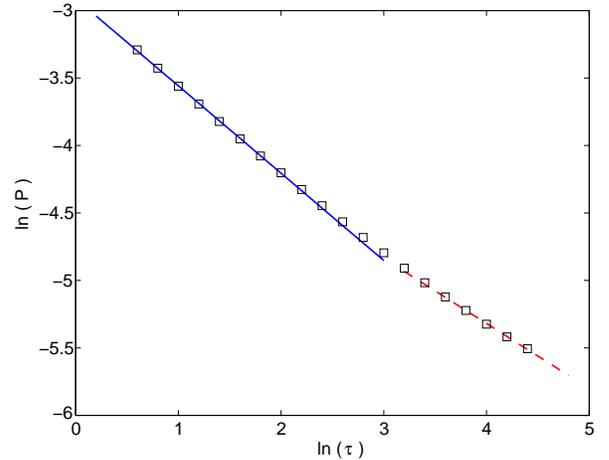}
\caption[]{Logarithmic plot of $P$ versus $\tau$ for $h_n = (t/\tau)
[0.2 \cos(\pi n) + \cos (\pi n/2 + \pi/4)]$ showing a cross-over between
two power laws. The two linear fits
near the beginning and end of the plot have slopes of $-0.65$ (blue solid)
and $-0.48$ (red dashed) respectively.} \label{Ptau} \end{figure}

Next, we consider the situation in which the magnetic field $h_n$ is periodic 
in $n$, but has several Fourier components. (We continue to assume that the 
period of $h_n$ is even, with $h_n$ being positive on half the sites and 
negative on the other half). For instance, let us consider a field of
the form $h_n = h [a_1 \cos(\pi n) + a_2 \cos (\pi n/2 + \pi/4)]$, where
$a_1$, $a_2$ and $\phi$ are constants; we then vary $h$ as $t/\tau$. (In
order to to focus on the scaling with respect to the quenching rate, we have 
chosen the phase of the period 4 term to be $\pi/4$ such that $\sin (2\phi) 
= 1$). If the $a_i$'s are small, the perturbation theory described in Sec. II 
shows that there is a coupling between the low-energy states at $k = 
\pm \pi/2$ at first order in $a_1$ and at second order in $a_2$; these would 
lead to a total excitation probability scaling as $a_1^{1/2}/
\tau^{1/2}$ and $(a_2)^{2/3} / \tau^{2/3}$ respectively. 
Assuming that $a_1$ and $a_2$ are both non-zero, we expect that for very 
large values of $\tau$, the $1/\tau^{1/2}$ scaling would dominate over the 
$1/\tau^{2/3}$ scaling. But if $\tau$ is not too large, 
then the $1/\tau^{2/3}$ scaling could dominate if the values of $\tau$, $a_1$ 
and $a_2$ are such that $(a_2)^{2/3} / \tau^{2/3} 
\gg a_1^{1/2}/\tau^{1/2}$. In particular, if $(a_2)^{2/3}
\gg a_1^{1/2}$, we may expect to see a $1/\tau^{2/3}$ scaling over a 
range of values of $\tau$ before crossing over to a $1/\tau^{1/2}$ scaling for 
very large values of $\tau$. Figure \ref{Ptau} shows a logarithmic plot of $P$
versus $\tau$ for the case $h_n = (t/\tau) [0.2 \cos(\pi n) + \cos (\pi n/2 
+ \pi/4)]$ (corresponding to $a_1 = 0.2$ and $a_2 =1$), along 
with two linear fits near the beginning and end of the plot. The linear fits 
have slopes of $-0.65$ and $-0.48$ which are in good agreement with the 
values $-2/3$ and $-1/2$ respectively. 

Finally, we may ask what happens if $Q=\pi p/q$, where $q$ is very large
and the variation of the potential is very slow. 
(For instance, this would be useful if we were interested in studying the 
case of a quasiperiodic pattern of the field $h_n$, and we approached such a 
pattern by taking rational approximations of the form $p/q$ with $q$ becoming 
progressively large). Unfortunately, it seems to be rather difficult to 
study the case of very large $q$ either analytically or numerically. 
Perturbation theory up to $q$-th order in $h$ would not be reliable, since the
energies appearing in the denominator of Eq.~(\ref{pert2}), $\sin (\pi s/q)$,
come arbitrarily close to zero if $s$ is held fixed and $q \to \infty$.
Numerical calculations based on time evolution with the $2q$-dimensional
Hamiltonian $H_k$ in Eq.~(\ref{hk}), with $h=t/\tau$, also become difficult
if $q$ is very large since there would be a large number of levels which
come close to each other at different times; the time evolution would
therefore need to be done very accurately to go from $t = - \infty$ to
$t = \infty$. In this limit, further studies and alternate methods are called 
for.

\section{Quenching in a Tomonaga-Luttinger liquid}

We have studied quenching in a periodic system of free fermions; we now turn 
to the manner in which this physics becomes modified in the presence of 
interactions between fermions. Our arguments borrow from the quenching 
treatment 
of Ref. \onlinecite{degrandi} for interacting bosons in a periodic potential 
favoring one particle per potential minimum, adapting it to the situation of 
interacting fermions as well as generalizing to multiple boson occupancy.

Tracing back to our initial starting point, the spin Hamiltonian of 
Eq.~(\ref{xxh}), the addition of a nearest-neighbor $zz$ coupling of the form 
$J_z \si_n^z \si_{n+1}^z$ results in an interaction term in the fermionic 
system. Employing the Jordan-Wigner transformation described in Sec. II leads 
to the addition of a four-fermion interaction of the form $c_n^\dg c_n
c_{n+1}^\dg c_{n+1}$ in the transformed fermionic Hamiltonian of 
Eq.~(\ref{ham1}).

We can treat the effect of these interactions on the quenching process 
described in the previous sections as follows. The quenching involves tuning 
the strength of the periodic potential
$h$ from $-\infty$ to $+\infty$ as a function of time, and once again, the 
most important contribution to quenching dynamics comes from the gapless 
region at vanishing $h$. We can thus consider the low-energy physics of 
the fermionic system for small $h$, namely the 
effective Hamiltonian Eq.~(\ref{heff}) which consists of right-moving and 
left-moving linear modes at Fermi points $k=\pm\pi/2$, $\psi_R$ and $\psi_L$, 
respectively, and a coupling between them of order $h^q$ derived from $q$-th 
order perturbation. In addition, we now have interactions and these can be 
treated within the standard context of bosonization, where the right and left 
moving modes $\psi_{R/L} \sim \exp [i \sqrt{\pi} ~(\mp\sqrt{K} \Phi + \theta /
\sqrt{K})]$ are expressed in terms of the bosonic fields $\Phi$ and $\theta$
\cite{gogolin}. In terms of the scalar field $\Phi$, 
the deviation of the density of fermions $\rho$ from the average density 
$\rho_0$ is given by $\rho - \rho_0 = - (\sqrt{K/\pi}) \pa \Phi /\pa x$. The 
Luttinger parameter $K$ characterizes the strength of the interactions 
($K<(>)1$ for repulsive (attractive) interactions, $K=1$ in the absence of 
interactions.) For a mapping from the spin model, the procedure can be 
considered in the range $-1 < J_z \le 1$; the parameter $K$ is then given by
$K =\frac{\pi}{2 \cos^{-1} (-J_z)}$, so that $K$ goes from $\infty$ to $1/2$ 
as $J_z$ goes from $-1$ to 1. 

The linear modes at $k=\pm\pi/2$ and the short-range fermion interactions 
together provide the standard action describing a Tomonaga-Luttinger liquid 
(TLL):
\beq S ~=~ \frac{1}{2} ~\int \int dt dx ~\left[~ \left( \frac{\pa \Phi}{\pa
t} \right)^2 ~-~ \left( \frac{\pa \Phi}{\pa x} \right)^2 \right]. 
\label{boson1} \eeq
(Here, the velocity of the bosonic excitations has been set to unity.)
The contribution of the $quench$ term to the action has the sine-Gordon form
\beq S_m ~\sim~ \int \int dt dx ~h^q ~\cos (2 \sqrt{\pi K} \Phi), 
\label{boson2} \eeq
Compared to the standard cases of interacting fermions that result in forms
similar to Eq.~(\ref{boson1}) and (\ref{boson2}), we emphasize that these terms
have been derived as the low-energy description of a system having a higher 
periodicity $2q$, resulting in the key dependence of the mass term $S_m$ on 
$q$.

Using Eqs.~(\ref{boson1}) and (\ref{boson2}) as the starting point, we invoke
renormalization group (RG) arguments \cite{gogolin} 
to derive the scaling of defect formation due to the quench. 
By computing the correlation function of the operator $\cos (2 \sqrt{\pi K} 
\Phi)$ at two different space-time points, one can show that it has mass 
dimension $K$; let us denote the coefficient of this operator in the action by 
$\mu$ in general. The parameter $\mu$ effectively becomes 
dependent on the length scale $L$, satisfying the RG equation
$d\mu /d \ln L = (2-K) \mu$. Given the initial value of $\mu (a) 
= h^q$ at some microscopic length scale $a$ (this could be either the lattice 
spacing or the average distance between the particles in a continuum model), 
the solution of the RG equation is $\mu (L) = h^q (L/a)^{2-K}$. Assuming 
that $h^q \ll 1$, we see that $\mu (L)$ is of order 1 at a 
length scale given by $L=\xi$, where $(\xi/a) \sim 1/h^{q/(2-K)}$. 
The condition that $\mu (\xi) \sim 1$ implies that we have reached
a strong coupling (disordered) regime at the length scale $\xi$, i.e.,
$\xi$ is the correlation length of the theory. Since $h$ denotes the deviation
from the quantum critical point, the relation $\xi/a \sim 1/h^{q/(2-K)}$ 
implies that the correlation length exponent is given by $\nu = q/(2-K)$. As 
argued in the previous sections, if $h$ is quenched through the QCP at $h=0$ 
at a rate given by $1/\tau$, the KZ scaling form then implies that the 
density of defects will scale as $1/\tau^{d\nu/(z\nu +1)}$. Hence, 
substituting the value of $\nu$ in this scaling form, the density of defects
scales as 
\beq \rho_D\sim 1/\tau^{q/(q+2-K)}. \label{rhoD} \eeq
This scaling behavior can be derived more rigorously by trivially 
modifying the treatment in Ref. \onlinecite{degrandi} to include the $q$ 
dependence. For the non-interacting case, $K=1$, this form reproduces the 
earlier expression $1/\tau^{q/(q+1)}$. In the presence of interactions, we 
see that the power varies continuously with $K$. We note that the power law 
$1/\tau^{q/(q+2-K)}$ is only valid for $1/2 < K < 2$. The value $K=2$
corresponds to a Kosterlitz-Thouless transition. For $K> 2$, the cosine term
in the action is irrelevant. It has been argued in Ref. \onlinecite{degrandi}
that the probability of excitations then receives contributions from all 
modes, not just the low-energy modes near $k = \pm \pi/2$; hence the scaling 
law has the form $1/\tau$ regardless of the value of $q$.

These arguments can be used to generalize the results of Ref. 
\onlinecite{degrandi} to a multi-boson situation. In addition to fermionic 
systems, the TLL form can be equally well applied to one-dimensional systems 
of repulsively interacting bosons; the Luttinger parameter $K$ goes from 
$\infty$ to $1$ as the strength 
of the repulsive interaction is varied from zero to $\infty$.
In particular, when the interaction strength is infinitely large, the bosonic
system is equivalent to a system of non-interacting spinless fermions
\cite{lieb63} with $K=1$. In the continuum of interacting bosons described 
by the action in Eq.~(\ref{boson1}), 
one can introduce a spatially periodic potential of the form 
$V(x,t) = h(t) \cos (2\pi \rho x/q)$, where $q$ is an integer. For
very large values of $h(t)$, the ground state would have exactly $q$ particles
residing in each of the minima of this potential. (Therein lies the 
generalization of Ref. \onlinecite{degrandi} which assumes $q=1$.) To analyze 
a quench of the form $h = t/\tau$, we note that there is a one-to-one 
correspondence between the interacting system of bosons and that of the 
fermions studied above. Focusing on the lower energy physics, once again 
Eq.~(\ref{boson1}) describes a gapless TLL with a Fermi momentum given by 
$k_F = \pi \rho$ while Eq.~(\ref{boson2}) describes the contribution coming 
from a small, non-vanishing $h$. The subsequent analysis
above also goes through for the bosonic case except for a redefinition of the
interaction parameter. We thus conclude that the density of defects (which
corresponds to some of the potential wells having either more than or less 
than $q$ particles) scales as $1/\tau^{q/(q+2-K)}$ as in Eq.~(\ref{boson2}) 
if $K< 2$, and as $1/\tau$ if $K > 2$.

\section{Conclusion}

To summarize, we have studied the dynamics of a spin-1/2 chain in which 
the amplitude of a spatially periodic magnetic field $h \cos (Qn + \phi)$ is 
slowly varied in time so as to take the system across a quantum critical 
point. We have equivalently analyzed 
a system of spinless fermions with a spatially periodic
and time-dependent chemical potential. We find 
that a quenching rate of $1/\tau$ takes the system to a final state having
a density of defects that scales as $1/\tau^{q/(q+1)}$ for $Q=\pi/q$;
the power depends on the period of the magnetic field. We have shown this
by using perturbation theory to derive an effective Hamiltonian for pairs of 
states near zero energy; this Hamiltonian has a parameter varying as a power
of the time, where this power also depends on the period of the field.
We have confirmed our results numerically in a number of cases. If the 
magnetic field has several Fourier components, the power law corresponding 
to the smallest period is found to dominate for very small values of 
$1/\tau$ although one may find a cross-over at intermediate values of
$\tau$ depending on the amplitudes of the different Fourier components. 
If the spin Hamiltonian has additional terms which can be mapped 
to interactions between the fermions, the power varies continuously 
with the interaction strength. To obtain a better understanding of some 
features of the excitation probability in the spin chain problem, we have 
numerically studied a two-level system in which a term in the Hamiltonian 
varies as a power law in time. We find that the excitation probability in 
this problem has a complicated dependence on the power resembling the results 
obtained for the many-body spin or fermion systems.

Our predictions, of interest from the perspective of critical phenomena and 
out-of-equilibrium dynamics, can be investigated in various physical systems. 
The most immediate experimental realizations would perhaps be in the area of 
cold atoms or molecules trapped in effectively one-dimensional optical 
lattices \cite{coldatoms}, where tuning 
parameters is easy and quenching has become an extensive topic of study. 
Another area would be one that has been studied in great depth and well 
characterized, namely, various
physical systems described by spin chain physics. In this case, the only 
additional ingredient necessary would be to subject the chains to a magnetic 
field having a periodic modulation in space and whose strength could be 
dynamically tuned. As yet another instance, recent proposals have employed
effectively spinless $p$-wave paired fermionic superconducting wires in 
schemes for topological quantum computation requiring tuning between 
topological and non-topological regions in these wires. As described in 
our previous work\cite{wade}, periodic potentials in such systems offer a 
way of studying topological aspects. A systematic study of the quench
work presented here generalized to include anomalous terms due to 
superconductivity (already initiated in Ref. \onlinecite{wade}) would be 
useful for the proposed schemes and as
a study in of itself on quenches in topological systems.
 
Our results have revealed how mode coupling can retain the slow 
out-of-equilibrium dynamics expected of quenching through a QCP while giving 
it a richer, more complex structure. Here, 
the mode coupling was via fragmentation of the Brillouin zone by a 
periodic potential. Future avenues would involve more intricate mode 
coupling, for instance, in the presence of quasiperiodic
potentials, or in disordered systems, which completely break translation 
invariance and can undergo delocalization-localization transitions, or 
even in systems of weakly coupled isolated quantum 
states. Finally, our studies have all been in the zero temperature limit. 
Since the quantum critical regime extends up to some finite range of 
temperatures, we expect the results presented here 
to persist at temperatures which are much smaller than the energy gap at the 
initial time \cite{patane}; a rigorous finite-temperature study is in order.

\section*{Acknowledgments}

For financial support, M.T. thanks CSIR, India, D.S. thanks DST, India under 
Project No. SR/S2/JCB-44/2010, and W.D. and S.V. thank the NSF under the 
grant DMR 0644022-CAR.
\vspace*{1truecm}

\appendix

\section{Non-linear quenching in the Landau-Zener system}

In this Appendix, we will study a generalization of the Landau-Zener problem 
\cite{lz,majorana} in which the Hamiltonian varies non-linearly in time. 
We consider a two-level system evolving with a time-dependent Hamiltonian as
\bea i \frac{d}{dt} \left( \begin{array}{c}
u_1 \\
u_2 \end{array} \right) = \left( \begin{array}{cc}
|t|^q sgn (t) & b \\
b & - |t|^q sgn (t) \end{array} \right) \left( \begin{array}{c}
u_1 \\
u_2 \end{array} \right), \non \\
&& \label{nonlin} \eea
where $b$ is a constant, and $q$ is a positive integer. It is sufficient to 
study this problem for the case that $b$ is real and positive, since the 
Hamiltonian for any complex value of $b$ is related by a unitary 
transformation to a Hamiltonian in which $b$ is replaced by $|b|$. The reason 
for studying the Hamiltonian in Eq.~(\ref{nonlin}) is that it is related by 
a unitary transformation to the one in Eq.~(\ref{tds2}), with $b$ playing the
role of $2k' (\tau^q /f(q\phi))^{1/(q+1)}$. Note that we have taken the time 
dependence in (\ref{nonlin}) to be of the form $|t|^q sgn (t)$; this has been 
done so that the ground states for $t \to -\infty$ and $t \to \infty$ are 
given by $(u_1,u_2) = (1,0)$ and $(0,1)$ respectively, just as they are for 
the case of linear time variation ($q=1$).

Note that we have taken the time dependence in (\ref{nonlin}) to be of the 
form $|t|^q sgn (t)$, rather than $t^q$; the two 
expressions agree for $q$ odd, but not for $q$ even. We have chosen the form 
$|t|^q sgn (t)$ so that for all values of $q$, the ground states for $t \to 
-\infty$ and $t \to \infty$ are given by $(u_1,u_2) = (1,0)$ and $(0,1)$ 
respectively, just as they are for the case of linear time variation ($q=1$).

At $t=-\infty$, we begin with the ground state $(u_1,u_2) = (1,0)$. We then 
numerically evolve the wave function as in Eq.~(\ref{nonlin}) up to $t = 
\infty$; at that point, we find the excitation probability given by $p(b)=|u_1
(\infty)|^2$. For $b=0$, we expect $p=1$ since the Hamiltonian has no matrix 
element connecting the initial and final ground states; hence the system 
remains in the initial ground state (the sudden
limit). For $b \to \infty$, we expect $p=0$ since the instantaneous ground and
excited states remain well separated at all times; hence the system follows
the instantaneous ground state (the adiabatic limit).
For $q=1$, one has the exact expression $p(b) = \exp (-\pi b^2)$ \cite{lz}.
For $q \ge 2$, the exact expression for $p(b)$ is not known.

\begin{figure}[t] 
\ig[width=3in]{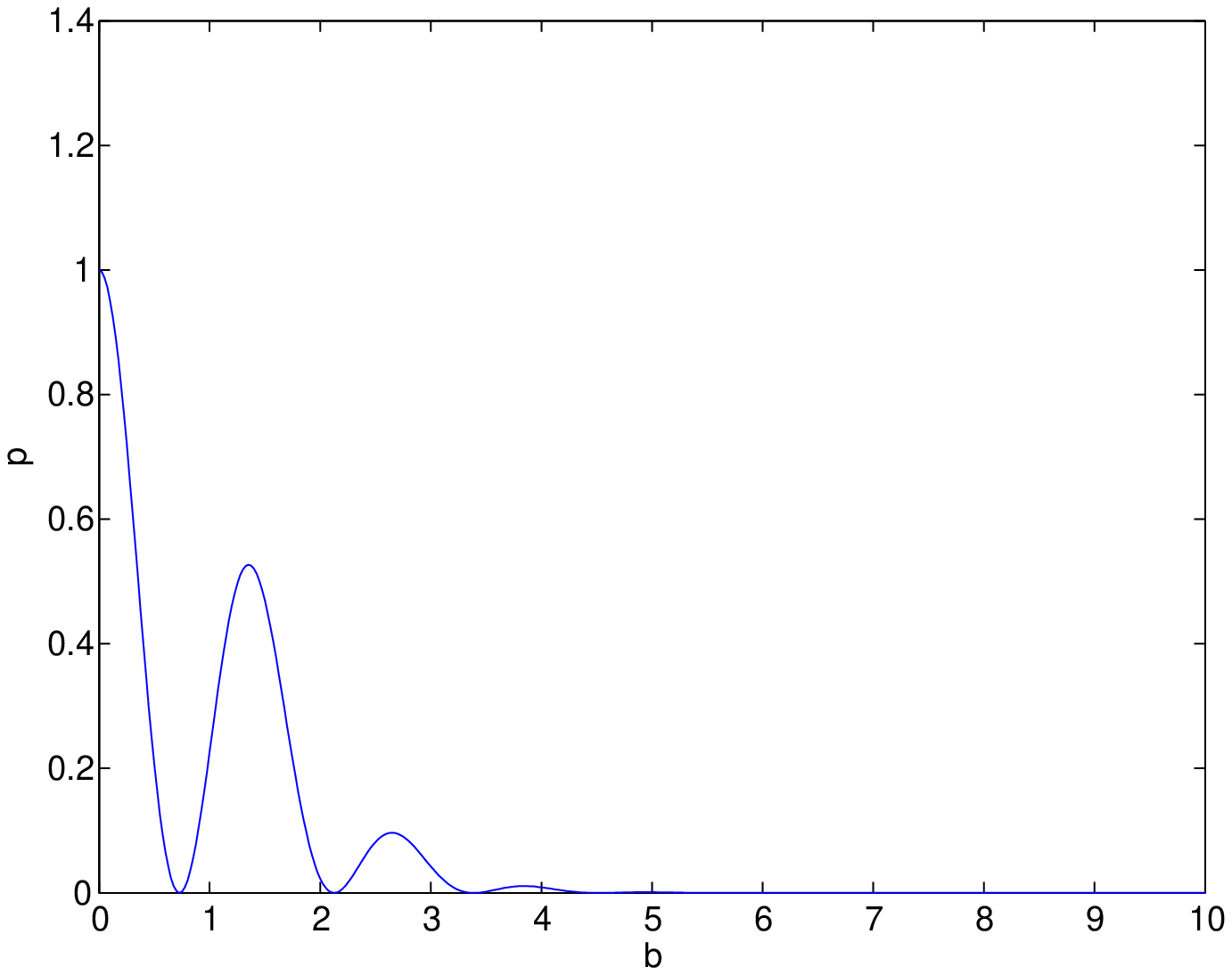}
\vspace*{.2in}
\ig[width=3in]{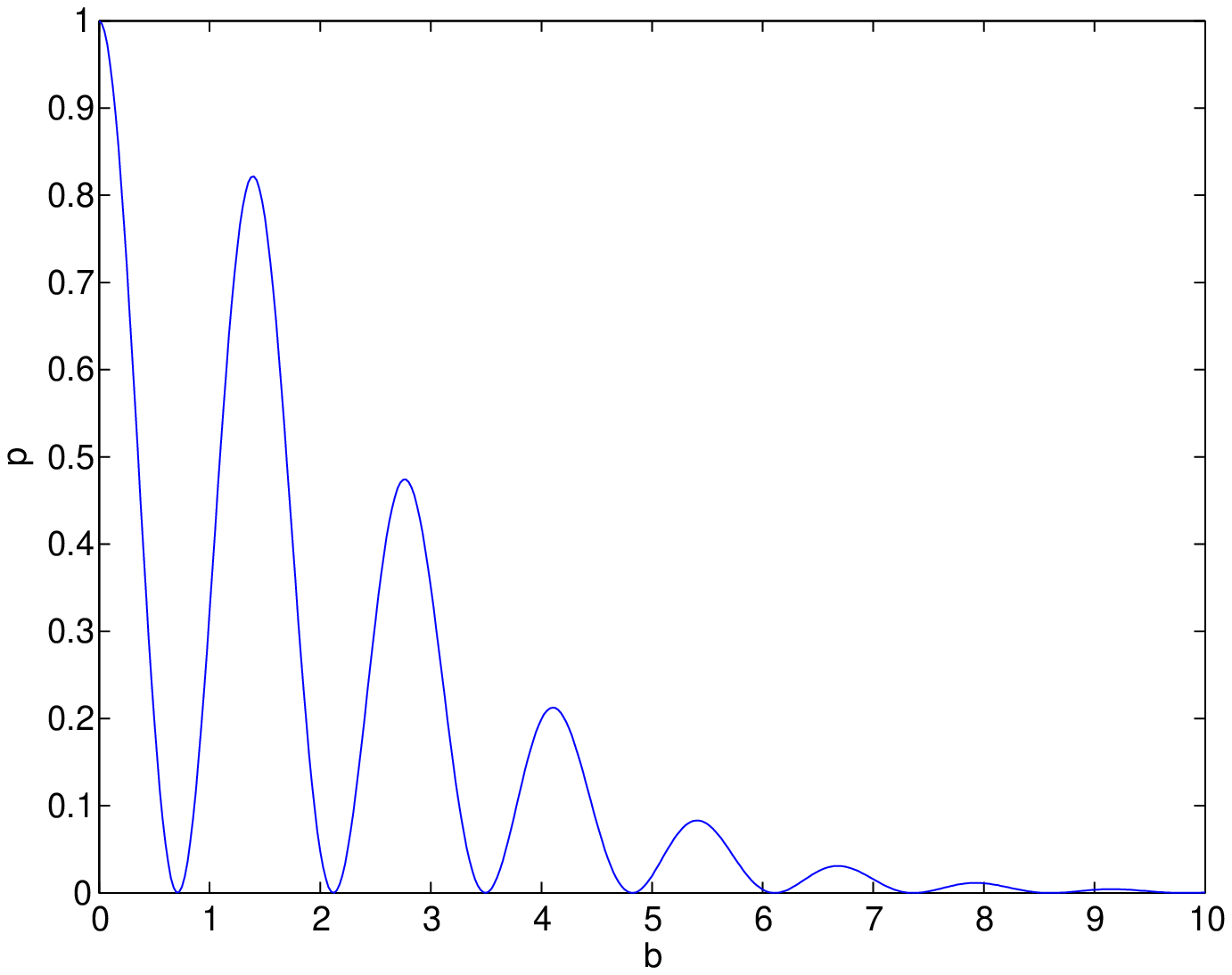}
\ig[width=3in]{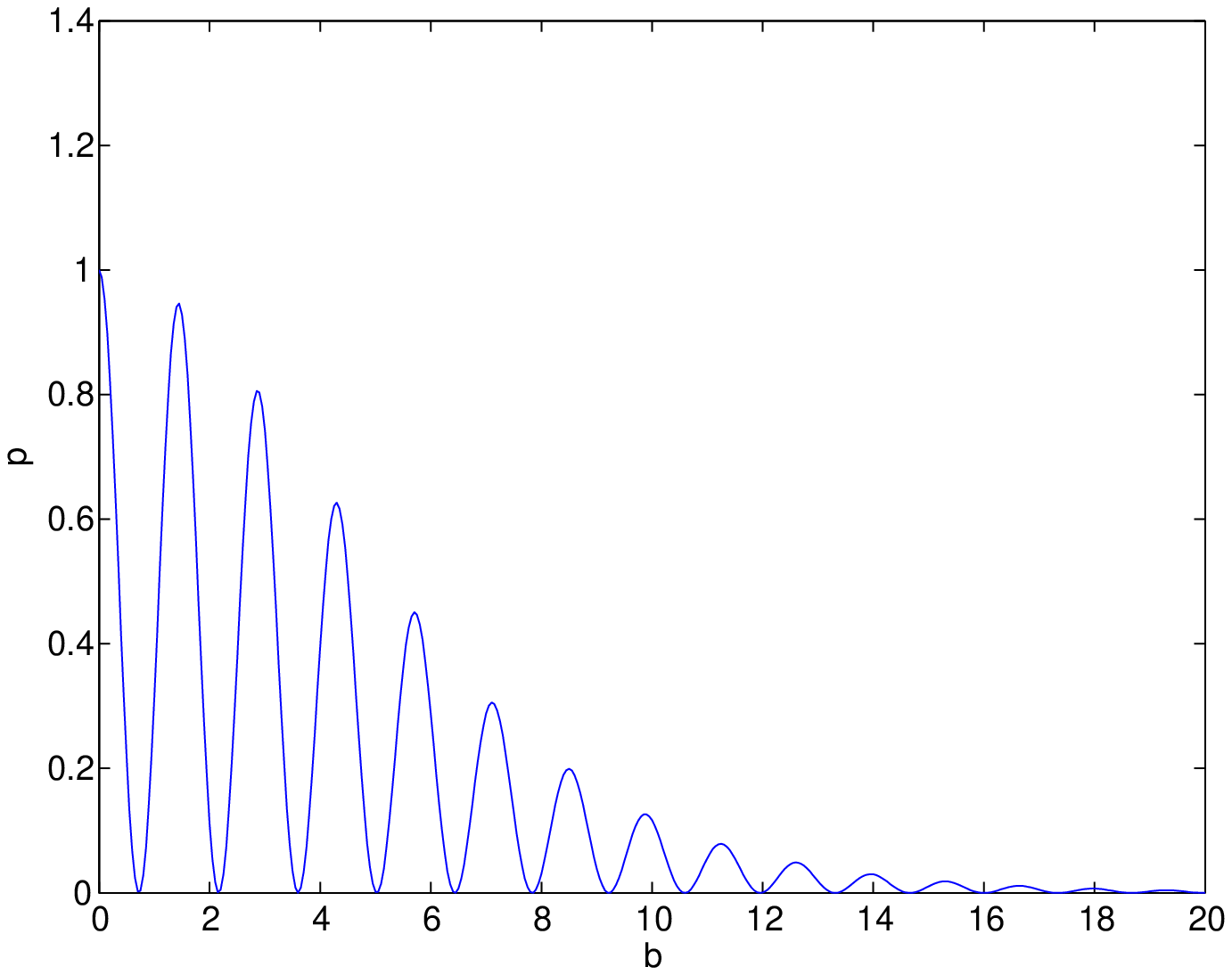}
\caption[]{Plot of $p$ versus $b$ for $q=5$ (top), $q=10$ (middle), and 
$q=20$ (bottom).} \label{pvsa} \end{figure}

Figure \ref{pvsa} shows how $p$ varies with $b$ for three different values of 
$q$. We see that $p(b)$ is a product of an oscillatory function and a decaying
function of $b$ (this seems to be true for all values of $q \ge 2$). The 
period of oscillations appears to approach $\pi /2$ for large values of $q$ 
and small values 
of $b$. We can understand this as follows. If $q \gg 1$, the function $|t|^q$ 
is much larger than 1 for $|t| > 1$ and much smaller than 1 for $|t|< 1$. Let 
us therefore make the approximation $|t|^q = \infty$ for $|t| > 1$ and $=0$ 
for $|t|< 1$. The Hamiltonian in Eq.~(\ref{nonlin}) can then be written as
\bea H &=& \left( \begin{array}{cc}
-\infty & b \\
b & \infty \end{array} \right) ~~~{\rm for}~~~ t ~<~ -1, \non \\
&=& \left( \begin{array}{cc}
0 & b \\
b & 0 \end{array} \right) ~~~{\rm for}~~~ -1 ~<~ t ~<~ 1, \non \\
&=& \left( \begin{array}{cc}
\infty & b \\
b & -\infty \end{array} \right) ~~~{\rm for}~~~ 1 ~<~ t. \eea
Since we begin with the wave function $(u_1,u_2) = (1,0)$ at $t=-\infty$,
and the energy levels are given by $\pm \infty$ from $t=-\infty$ up to $t=1$,
the wave function remains equal to $(1,0)$ (times a phase factor due to
time evolution with infinite energy) up to $t=-1$. Then the Hamiltonian is 
independent of time from $t=-1$ to $t=1$; this is known as the waiting problem
\cite{divakaran}. It is easy to solve for the time evolution during
this period; we find that the wave function at $t=1$ is given by
$(\cos (2b), i\sin (2b))$ times a phase factor. From $t=-1$ to $t=\infty$,
the energy levels are again given by $\pm \infty$; hence the wave function
at $t=\infty$ is given by $u_1 = \cos (2b)$ times a phase factor and $u_2=
\sin (2b))$ times another phase factor. We thus obtain $p(b)=\cos^2 (2b)$. 
This explains the oscillations in $p$ versus $b$ with period $\pi /2$ and 
amplitude equal to $1/2$ for large values of $q$ and small values of $b$. 
However, this simple explanation breaks down for large values of $b$ where 
the period of oscillations starts deviating from $\pi /2$ and the amplitude of
oscillations goes to zero as we can see in Fig.~\ref{pvsa}.

\begin{figure}[t] \ig[width=3in]{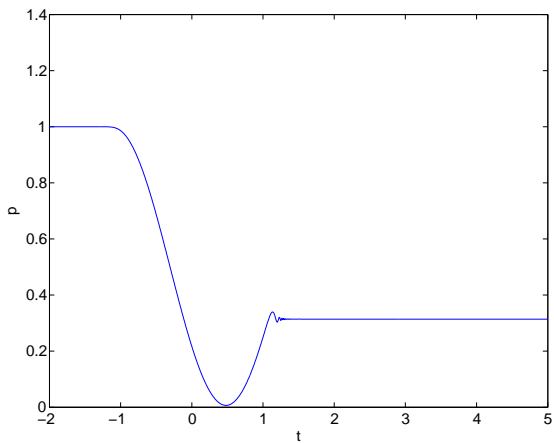}
\caption[]{Plot of $p$ versus $t$ for $q=20$ and $b=1$.} \label{pvst} 
\end{figure}

\begin{figure}[t] 
\vspace*{.2in}
\ig[width=3in]{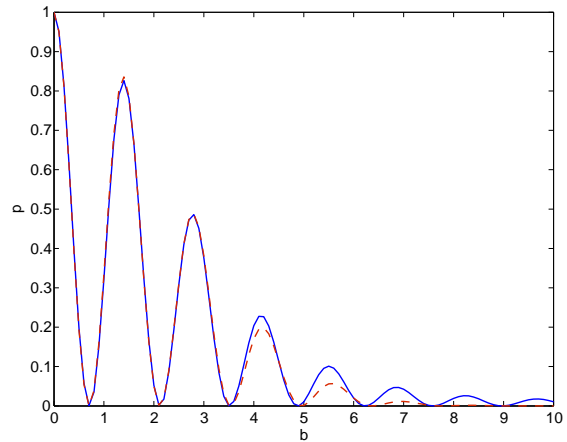}
\caption[]{Two-parameter fit to plot of $p$ versus $b$ for $q=10$. The fitting
function $p(b) = e^{-c_1 b^2} \cos^2 (2c_2 b)$ gives $c_1 =0.091$ and $c_2 =
1.11$.} \label{fitpvsa} \end{figure}

Figure \ref{pvst} confirms the 
scenario presented above for the variation of the
excitation probability $p = |u_1|^2$ versus the time $t$. Namely, $p$ does not
change much before $t=-1$ or after $t=1$, but it oscillates between $t=-1$ and 
$1$. Figure \ref{fitpvsa} shows a two-parameter fit to a plot of $p$ versus $b$
for $q=10$; a function of the form $p(b) = e^{-c_1 b^2} \cos^2 (2c_2 b)$ with 
$c_1 =0.091$ and $c_2 =1.11$ gives a good fit for $b \lesssim 3.5$, but the fit
becomes progressively worse for larger values of $b$. For $q=20$, we find that
the same function of $p$ versus $b$ gives a good fit with $c_1 =0.020$ and
$c_2 =1.08$ for $b \lesssim 5.5$. Note that our analysis above predicts that 
$c_2$ should be equal to 1.



The study of the two-level system in this Appendix sheds some light on 
the results we obtained in Sec. III for the spin system. In particular, the 
observation in Fig.~\ref{pvsa} that as $q$ increases, $p(b)$ shows 
increased oscillation as a function of $b$ (while the amplitude of the 
oscillations goes to zero for large $b$) explains similar features seen 
in Figs.~\ref{pkq3} and \ref{pkq11} for the spin system.

\end{document}